\documentstyle[11pt]{article}
\bibliography{plain}
\title{\bf Concurrence as a Relative Entropy with Hilbert-Schmidt Distance
in Bell Decomposable States } \vspace{20mm}
\author{S. J. Akhtarshenas  $^{a,b,c}$ \thanks{E-mail:akhtarshenas@tabrizu.ac.ir}
 , M. A. Jafarizadeh$^{a,b,c}$ \thanks{E-mail:jafarizadeh@tabrizu.ac.ir}
\\
\\
$^a${\small Department of Theoretical Physics and Astrophysics,
 Tabriz University, Tabriz 51664, Iran.} \\
$^b${\small Institute for Studies in Theoretical Physics and Mathematics,
 Tehran 19395-1795, Iran.} \\
$^c${\small Research Institute for Fundamental Sciences, Tabriz
51664, Iran.}} \pagebreak

\begin{document}
\maketitle \vspace{15mm}
\newpage
\begin{abstract}
Hilbert-Schmidt distance reduces to Euclidean distance in Bell
decomposable states. Based on this, entanglement of these states
are obtained according to the protocol proposed in Ref. [V. Vedral
et al, Phys. Rev. Lett. {\bf 78}, 2275 (1995)] with
Hilbert-Schmidt distance. It is shown that this measure is equal
to the concurrence and thus can be used to generate entanglement
of formation. We also introduce a new measure of distance and show
that under the action of restricted LQCC operations, the
associated measure of entanglement transforms in the same way as
the concurrence transforms .


{\bf PACs Index: 03.65.Ud  }

\end{abstract}
\pagebreak

\vspace{7cm}

\section{Introduction}
Perhaps, quantum entanglement is the most non classical features
of quantum mechanics \cite{EPR,shcro} which has recently been
attracted much attention. It plays a central role in quantum
information theory \cite{ben1,ben2,ben3}. Entanglement is usually
arise from quantum correlations between separated subsystems which
can not be created by local actions on each subsystem. By
definition, a mixed state $\rho$ of a bipartite system is said to
be separable (non entangled) if it can be written as a convex
combination of pure product states
\begin{equation}
\rho=\sum_{i}p_{i}\left|\phi_{i}^{A}\right>\left<\phi_{i}^{A}\right|
\otimes\left|\psi_{i}^{B}\right>\left<\psi_{i}^{B}\right|,
\end{equation}
where $\left|\phi_{i}^{A}\right>$ and $\left|\psi_{i}^{B}\right>$
are pure states of subsystems $A$ and $B$, respectively. Although,
in a  pure state of bipartite systems it is easy to check whether
a given state is, or is not entangled, the question is yet an open
problem in the case of mixed states.

There is also an increasing attention in quantifying entanglement,
particularly for mixed states of a bipartite system, and a number
of measures have been proposed \cite{ben3,ved1,ved2,woot}. Among
them the entanglement of formation has more importance, since it
intends to quantify the resources needed to create a given
entangled state. Vedral et al. in \cite{ved1,ved2} introduced a
class of distance measures suitable for entanglement measures.
They also showed that the quantum relative entropy and the Bures
metric satisfy three conditions that a good measure of
entanglement must satisfy and can therefore be used as generators
of measures of entanglement.

Hilbert-Schmidt distance have been used as a measure of distance
in \cite{witte}.  They obtained the entanglement of Bell
decomposable states and part of pure states for  $2\otimes 2$
systems according to H-S distances.

In this paper we show that H-S distance reduces to Euclidean
distance for special kind of $2\otimes 2$ states, called Bell
decomposable states. Based on this, we can rather easily calculate
entanglement measure associated with H-S distance of these states.
We also show that thus obtained quantity is equal to the
concurrence and thus can be used to generate entanglement of
formation.

Finally, we present a new measure of distance in operators space
and show that the corresponding entanglement measure reduces to
concurrence. Starting from BD sates, we perform quantum operations
and classical communications (LQCC) \cite{lind,kent} and as
consequence one obtains new entangled mixed density matrices with
entanglement measure with a functionality analogous to the
concurrence.

The paper is organized as follows. In section 2 we review BD
states and present a perspective of their geometry. In section 3
we show that H-S distance of these states reduces to Euclidean
distance. In section 4 we give a brief review of Wootters'
concurrence, then we evaluate entanglement measure associated with
H-S distance and show that it is equal to the concurrence. In
section 5 we introduce a new measure of distance and show that its
corresponding entanglement measure is also equal to the
concurrence for BD states. We perform LQCC action on these states
and show that  under LQCC, the corresponding entanglement measure
transforms in the same way as the concurrence transforms. The
paper is ended with a brief conclusion.

\section{Bell Decomposable States}
In this section we briefly review Bell decomposable (BD) states
and some of their properties. A BD state is defined by
\begin{equation}
\rho=\sum_{i=1}^{4}p_{i}\left|\psi_i\right>\left<\psi_i\right|,\quad\quad
0\leq p_i\leq 1,\quad \sum_{i=1}^{4}p_i=1,
 \label{BDS1}
\end{equation}
where $\left|\psi_i\right>$ is Bell state given by
\begin{eqnarray}
\label{BS1} \left|\psi_1\right>=\left|\phi^{+}\right>
=\frac{1}{\sqrt{2}}(\left|\uparrow\uparrow\right>+
\left|\downarrow\downarrow\right>), \\
\label{BS2}\left|\psi_2\right>=\left|\phi^{-}\right>
=\frac{1}{\sqrt{2}}(\left|\uparrow\uparrow\right>-
\left|\downarrow\downarrow\right>), \\
\label{BS3}\left|\psi_3\right>=\left|\psi^{+}\right>
=\frac{1}{\sqrt{2}}(\left|\uparrow\downarrow\right>+
\left|\downarrow\uparrow\right>), \\
\label{BS4}\left|\psi_4\right>=\left|\psi^{-}\right>
=\frac{1}{\sqrt{2}}(\left|\uparrow\downarrow\right>-
\left|\downarrow\uparrow\right>).
\end{eqnarray}
In terms of Pauli's matrices, $\rho$ can be written as

\begin{equation}
\rho=\frac{1}{4}(I\otimes I+\sum_{i=1}^{3}
t_i\sigma_{i}\otimes\sigma_{i}), \label{BDS2}
\end{equation}
where

\begin{equation}\label{t-p}
\begin{array}{rl}
t_1=&p_1-p_2+p_3-p_4,  \\
t_2=&-p_1+p_2+p_3-p_4, \\
t_3=&p_1+p_2-p_3-p_4.
\end{array}
\end{equation}

From positivity of $\rho$ we get
\begin{equation}\label{T1}
\begin{array}{rl}
1+t_1-t_2+t_3\geq & 0,  \\
1-t_1+t_2+t_3\geq & 0,  \\
1+t_1+t_2-t_3\geq & 0,  \\
1-t_1-t_2-t_3\geq & 0.
\end{array}
\end{equation}
These equations form a tetrahedral  with its vertices located at
$(1,-1,1)$, $(-1,1,1)$, $(1,1,-1)$, $(-1,-1,-1)$ \cite{horo2}. In
fact these vertices are Bell states given in Eqs. (\ref{BS1}) to
(\ref{BS4}), respectively.

According to the Peres and Horodecki's condition for separability
\cite{peres,horo1}, a 2-qubit state is separable if and only if
its partial transpose is positive. This implies that $\rho$ given
in Eq. (\ref{BDS2}) is separable if and only if $t_i$ satisfy Eqs.
(\ref{T1}) and
\begin{equation}\label{T2}
\begin{array}{rl}
1+t_1+t_2+t_3\geq & 0,  \\
1-t_1-t_2+t_3\geq & 0,  \\
1+t_1-t_2-t_3\geq & 0,  \\
1-t_1+t_2-t_3\geq & 0.
\end{array}
\end{equation}

Inequalities (\ref{T1}) and (\ref{T2}) form an octahedral with its
vertices located at $O_1^{\pm}=(\pm 1,0,0)$, $O_2^{\pm}=(0,\pm
1,0)$ and $O_3^{\pm}=(0,0,\pm 1)$. Hence, tetrahedral of Eqs.
(\ref{T1}) is divided into five regions. Central regions, defined
by octahedral, are separable states. There are also four smaller
equivalent tetrahedral corresponding to entangled states. Each
tetrahedral takes one Bell state as one of its vertices. Three
other vertices of each tetrahedral form a triangle which is its
common face with the octahedral (See Fig. 1).

\section{Hilbert-Schmidt distance for BD states}

In the Hilbert-Schmidt (H-S)  space density matrices are regarded
as vectors rather than operators in the conventional quantum
mechanics. The inner product between two operators $A$ and $B$ in
H-S space is defined as
\begin{equation}
\left<A,B\right>=tr(A^\dag B),
\end{equation}
where $A$ and $B$ are $4\times4$ matrices.

Using definition of inner product, we can define the norm of a
given vector $A$ in H-S space as
\begin{equation}
\|A\|=\sqrt{\left<A,A\right>}=\sqrt{tr(A^\dag A)},
\end{equation}
which is called trace norm. Hence it is natural to obtain the
distance between two vectors $A$ and $B$ in H-S space as
\begin{equation}
d=\|A-B\|.
\end{equation}

Now, we show that for BD states H-S distance is equivalent to
Euclidean distance. Let us consider two density matrix $\rho$ and
$\rho^\prime$. Using Eq. (\ref{BDS2}) we can expand them in terms
of Pauli's matrices
\begin{eqnarray}
\rho=\frac{1}{4}(I\otimes I+\sum_{i=1}^{3}
t_i\sigma_{i}\otimes\sigma_{i}), \\
\rho^\prime=\frac{1}{4}(I\otimes I+\sum_{i=1}^{3} t_i^{\prime}
\sigma_{i}\otimes\sigma_{i}). \label{BDS3}
\end{eqnarray}
Using the above equations one can straightforwardly evaluate their
H-S distance, where we have
\begin{equation}
\|\rho-\rho^{\prime}\|=\sqrt{tr(\rho-\rho{\prime})^2}=
\frac{1}{2}\sqrt{\sum_{i=1}^{3}(t_i-t_i^{\prime})^2}. \label{euc}
\end{equation}

Now we can to use the above results to obtain concurrence in the
next section.
\section{Concurrence as relative entropy with H-S distance}
From the various measures proposed to quantify entanglement, the
entanglement of formation has a special position which in fact
intends to quantify the resources needed to create a given
entangled state \cite{ben3}. Wootters in \cite{woot} has shown
that for a 2-qubit system entanglement of formation of a mixed
state $\rho$ can be defined as
\begin{equation}
E(\rho)=H(\frac{1}{2}+\frac{1}{2}\sqrt{1-C^2}),
\end{equation}
where $H(x)=-x\ln{x}-(1-x)\ln{(1-x)}$ is binary entropy and
$C(\rho)$, called concurrence, is defined by
\begin{equation}
C(\rho)=\max\{0,\lambda_1-\lambda_2-\lambda_3-\lambda_4\},
\end{equation}
where the $\lambda_i$ are the non-negative eigenvalues, in
decreasing order, of the Hermitian matrix
$R\equiv\sqrt{\sqrt{\rho}{\tilde \rho}\sqrt{\rho}}$ and
\begin{equation}\label{rhotilde}
{\tilde \rho}
=(\sigma_y\otimes\sigma_y)\rho^{\ast}(\sigma_y\otimes\sigma_y),
\end{equation}
where $\rho^{\ast}$ is the complex conjugate of $\rho$ when it is
expressed in a fixed basis such as $\{\left|\uparrow\right>,
\left|\downarrow\right>\}$, and $\sigma_y$ is
$\left(\begin{array}{cc}0 & -i \\ i & 0
\end{array}\right)$ on the same basis.

Now, let us consider a Bell decomposable state given by
(\ref{BDS2}). One can show that for these states ${\tilde
\rho}=\rho$ and thus $R=\sqrt{\sqrt{\rho}{\tilde
\rho}\sqrt{\rho}}=\rho$. Calculating the eigenvalues of $R$ we get
\begin{equation}\label{wotcon}
C=\max\{0,-\frac{1}{2}(1+t_1+t_2+t_3)\}=\max\{0,2p_4-1\}.
\end{equation}
In the sequel we obtain concurrence given in Eq. (\ref{wotcon})
from an entirely different approach. Vedral et al. in
\cite{ved1,ved2} introduced a class of distance measures suitable
for entanglement measures. According to their methods,
entanglement measure for a given state $\rho$ is defined as
\begin{equation}\label{D}
E(\rho)=\min_{\sigma\in {\mathcal D}} D(\rho\parallel\sigma),
\end{equation}
where $D$ is any measure of distance (not necessarily a metric)
between two density matrix $\rho$ and $\sigma$ and ${\mathcal D}$
is the set of all separable states. They have also shown that
quantum entropy and Bures metric satisfy three conditions that a
good measure of entanglement must satisfy and can therefore be
used as generators of measures of entanglement.

Witte et al. used H-S distance as a candidate for Eq. (\ref{D}).
They obtained entanglement of the BD states based on H-S distance
from a rigorous method.

 Now entanglement of BD states can be easily evaluated by using
 Eq. (\ref{euc}), where H-S distance is equal to Euclidean
 distance.

Let us consider state $\rho$ in the entangled tetrahedral
corresponding to singlet state ($p_4\geq\frac{1}{2}$). It can be
easily seen that the nearest separable surface to this state is
$x_1+x_2+x_3+1$  which is its common face with octahedral  (See
Fig. 2). If $\rho_s$ denotes nearest separable density matrix to
$\rho$, then it must lie on this separable surface. This means
that $p_4^\prime=\frac{1}{2}$. Minimizing the Euclidean distance
between $\rho$ and $\rho_{s}$  with the constraints 
$p_4^\prime=\frac{1}{2}$ and $\sum_{i=1}^{4}p_i^\prime=1$, we get
\begin{eqnarray}\label{pp}
p_i^{\prime}=p_i+\frac{1}{3}\left(p_4-\frac{1}{2}\right)\quad\quad{\mbox
for} \quad i=1,2,3 \qquad  \mbox{and}\quad
p_4^{\prime}=\frac{1}{2}.
\end{eqnarray}

In terms of parameters $t_i$, Eq. (\ref{pp}) takes the following
form
\begin{equation}\label{tp}
t_i^{\prime}=t_i-\frac{1+t_1+t_2+t_3}{3}.
\end{equation}

Using the above result and Eq. (\ref{euc}) we obtain
\begin{equation}\label{HSdis}
D(\rho\parallel\sigma)=-\frac{1+t_1+t_2+t_3}{2\sqrt{3}}=\frac{C}{\sqrt{3}},
\end{equation}
Now we can define entanglement of $\rho$ as
\begin{equation}\label{HScon}
E(\rho)=\sqrt{3}D(\rho\parallel\sigma)=C
\end{equation}
Right hand side of Eq. (\ref{HScon}) is the concurrence given in
Eq. (\ref{wotcon}).

\section{Tilde norm }
Here in this section, we introduce a new norm defined as
\begin{equation}\label{newnorm}
\widetilde{\|A \|}:=\sqrt{tr(A\,{\tilde A} )},
\end{equation}
where ${\tilde A}$ is defined according to Eq. (\ref{rhotilde}).
With respect to this norm the distance between two density
matrices $\rho_1$ and $\rho_2$ is defined by
\begin{equation}\label{newdis}
d=\widetilde{\|\rho_1-\rho_2\|}=\sqrt{tr((\rho_1-\rho_2)({\tilde
\rho}_1-{\tilde \rho}_2))}.
\end{equation}
In the sequel we use Eq. (\ref{newdis}) as a measure of distance.
We define entanglement  of a state $\rho$ by
\begin{equation}\label{newent}
E(\rho):=\min_{\sigma\in{\cal D}}\widetilde{\|\rho-\sigma\|}.
\end{equation}
It is straightforward  to see that for BD states the above
distance reduces to H-S distance. Hence the separable state
$\sigma$ which minimize expression (\ref{newent}) is the same as
$\rho_{s}$ given in Eq. (\ref{tp}).

In the sequel we perform local quantum operations and classical
communications (LQCC) on BD states to study the change of
entanglement.

A general LQCC transformation is defined by
\begin{equation}\label{lqcc}
\rho^{\prime}=\frac{(A\otimes B)\rho(A\otimes
B)^{\dag}}{tr((A\otimes B)\rho(A\otimes B)^{\dag})},
\end{equation}
where operators $A$ and $B$ can be written as
\begin{equation}
A\otimes B=U_{A}\,f^{\mu,a,{\bf m}}\otimes U_{B}\,f^{\nu,b,{\bf
n}},
\end{equation}
where $U_{A}$ and $U_{B}$ are unitary operators acting on
subsystems $A$ and $B$, respectively and the filters
$f^{\mu,a,{\bf m}}$ and $f^{\nu,b,{\bf n}}$ are defined by
\begin{equation}\label{filt}
\begin{array}{rl}
f^{\mu,a,{\bf m}}= & \mu(I_2 + a\,{\bf m}.{\bf \sigma}), \\
f^{\nu,b,{\bf n}}= & \nu(I_2 + b\,{\bf n}.{\bf \sigma}).
\end{array}
\end{equation}
As it is shown in Refs. \cite{lind,kent}, the concurrence of the
state $\rho$ transforms under LQCC of the form given in Eq.
(\ref{lqcc}) as
\begin{equation}\label{conlqcc}
C(\rho^{\prime})=\frac{\mu^2\,\nu^2(1-a^2)(1-b^2)}{t(\rho;\mu,a,{\bf
m},\nu,b,{\bf n})}\,C(\rho),
\end{equation}
where $t(\rho;\mu,a,{\bf m},\nu,b,{\bf m})=tr((A\otimes
B)\rho(A\otimes B)^{\dag})$.

Now we perform LQCC transformation on states $\rho$ and $\rho_{s}$
\begin{eqnarray}\label{lqcc2}
\rho^{\prime}=\frac{U_{A}\,f^{\mu,a,{\bf m}}\otimes
U_{B}\,f^{\nu,b,{\bf n}}\,\,\rho\,\,f^{\mu,a,{\bf
m}}\,U_{A}^{\dag}\otimes f^{\nu,b,{\bf
n}}\,U_{B}^{\dag}}{t(\rho;\mu,a,{\bf m},\nu,b,{\bf
n})}, \\
\rho_{s}^{\prime}=\frac{U_{A}\,f^{\mu,a,{\bf m}}\otimes
U_{B}\,f^{\nu,b,{\bf n}}\,\,\rho_{s}\,\,f^{\mu,a,{\bf
m}}\,U_{A}^{\dag}\otimes f^{\nu,b,{\bf
n}}\,U_{B}^{\dag}}{t(\rho_{s};\mu,a,{\bf m},\nu,b,{\bf n})}.
\end{eqnarray}

Under LQCC transformation ${\tilde \rho}$ and ${\tilde \rho_{s}}$
change as \cite{lind}
\begin{eqnarray}\label{lqcc3}
{\tilde \rho}^{\prime}=\frac{U_{A}\,f^{\mu,a,-{\bf m}}\otimes
U_{B}\,f^{\nu,b,-{\bf n}}\,\,{\tilde \rho}\,\,f^{\mu,a,-{\bf
m}}\,U_{A}^{\dag}\otimes f^{\nu,b,-{\bf
n}}\,U_{B}^{\dag}}{t(\rho;\mu,a,{\bf m},\nu,b,{\bf
n})}, \\
{\tilde \rho}_{s}^{\prime}=\frac{U_{A}\,f^{\mu,a,-{\bf m}}\otimes
U_{B}\,f^{\nu,b,-{\bf n}}\,\,{\tilde \rho}_{s}\,\,f^{\mu,a,-{\bf
m}}\,U_{A}^{\dag}\otimes f^{\nu,b,-{\bf
n}}\,U_{B}^{\dag}}{t(\rho_{s};\mu,a,{\bf m},\nu,b,{\bf n})}.
\end{eqnarray}

 Now, taking into account that $f^{\mu,a,{\bf m}}\,f^{\mu,a,-{\bf m}}=\mu^2(1-a^2)I_2$
  and $f^{\nu,b,{\bf n}}\,f^{\nu,b,-{\bf n}}=\nu^2(1-b^2)I_2$, and
  using Eq. (\ref{newdis}) we can evaluate the  distance between two
states $\rho^{\prime}$ and $\rho_{s}^{\prime}$, where we have
\begin{equation}
\|\rho^{\prime}-\rho_{s}^{\prime}\|=\mu^4\,\nu^4\,(1-a^2)^2\,(1-b^2)^2
tr\left(\left(\frac{\rho}{t(\rho)}-\frac{\rho_{s}}{t(\rho_s}\right)
\left(\frac{{\tilde \rho}}{t(\rho)}-\frac{{\tilde
\rho}_{s}}{t(\rho_s)}\right)\right)
\end{equation}

where $t(\rho)$ and $t(\rho_s)$ are defined according to Eqs.
(\ref{BDS2})and (\ref{tp}) as
\begin{equation}\label{trho}
\begin{array}{rl}
t(\rho)= t(\rho;\mu,a,{\bf m},\nu,b,{\bf n})=& \mu^2\,\nu^2
\left((1+a^2)(1+b^2)+4ab\sum_{i=1}^{3}m_i\,t_i\,n_i\right) \\
t(\rho_s)= t(\rho_s;\mu,a,{\bf m},\nu,b,{\bf n})=& \mu^2\,\nu^2
\left((1+a^2)(1+b^2)+4ab\sum_{i=1}^{3}m_i\,t_i^{\prime}\,n_i\right) \\
 =& t(\rho)-\frac{4}{3}\mu^2\,\nu^2\,a\,b\,{\bf m}.{\bf
 n}\,(1+t_1+t_2+t_3) \\
=& t(\rho)-\frac{2}{3}\mu^2\,\nu^2\,a\,b\,{\bf m}.{\bf
 n}\,C,
\end{array}
\end{equation}
where in the last line we used concurrence $C$ given in
(\ref{wotcon}). In the sequel we make special choices for LQCC
transformations. Let us consider cases that either ${\bf m}.{\bf
n}=0$ or $a\,b=0$, that is, $t(\rho)=t(\rho_s)$. In this cases we
get
\begin{equation}\label{ccp}
E(\rho^{\prime})=\frac{\mu^2\,\nu^2\,(1-a^2)\,(1-b^2)}
{t(\rho;\mu,a,{\bf m},\nu,b,{\bf m})}\,E(\rho).
\end{equation}
Comparison of the above result with Eq. (\ref{conlqcc}) shows that
under the action of LQCC, the newly defined measure of
entanglement changes in the same way  as the concurrence changes,
therefore it is the same as the concurrence.

{\large \bf{Conclusion}}\\ We have  shown that H-S distance is
equivalent to Euclidean distance for BD sates. Based on this, H-S
entanglement measure of these states are easily obtained and it
has been  shown that, H-S entanglement measure  is equal to the
concurrence. Based on spin flipping transformation of  density
matrices, we have introduced a new measure of distance, and we
have shown that its corresponding measure of entanglement is equal
to the concurrence for BD states. Starting from BD states together
by performing restricted LQCC action, we have shown that the
transformed entanglement measure is equal to the concurrence.

\newpage

\vspace{10mm}

{\Large {\bf Figure Captions}}

\vspace{10mm} Figure 1: All BD states correspond to the interior
points of  tetrahedral. Vertices $P_{1}$, $P_{2}$, $P_{3}$ and
$P_{4}$ denote projectors corresponding to Bell states defined by
Eqs. (\ref{BS1}) to (\ref{BS4}), respectively. The interior points
of octahedral correspond to separable states.

\vspace{10mm}

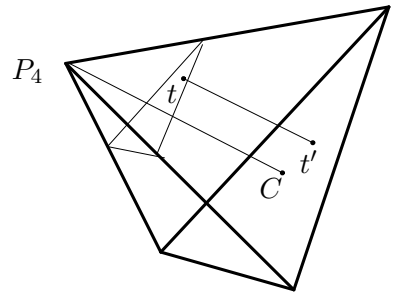
Figure 2: Entangled tetrahedral corresponding to singlet state.
The points of line $P_4\,C$ correspond to  entangled Werner
states. Points $t$ and $t^{\prime}$ correspond to a generic BD
state $\rho$ and associated nearest separable state $\rho_{s}$,
respectively.

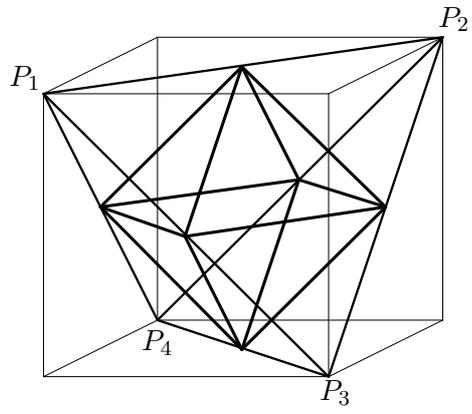
\end{document}